\documentclass[12pt]{elsarticle}
\usepackage{graphicx}
\usepackage{amsmath}
 \usepackage{subfig}

\journal{Mechatronics}

\begin{document}

\title{Computed-torque method for the control of a 2 DOF orthosis actuated through pneumatic artificial muscles: a specific case for the rehabilitation of the lower limb.}

\author{Flavio Prattico*}  
\address{Dipartimento di Ingegneria Industriale e dell'Informazione e di Economia, Universit\`a degli studi dell'Aquila, Italy\\
*Corresponding Author, flavioprattico@gmail.com, via G. Gronchi 18, 67100, L'Aquila, Italy}
\author{Mohd Azuwan Mat Dzahir}
\address{Bio-Science and Engineering Department, Shibaura Institute of Technology, Japan}
\author{Shin-ichiroh Yamamoto}
\address{Bio-Science and Engineering Department, Shibaura Institute of Technology, Japan}

\begin{abstract}
In this paper we give a new control model based on the so called computed-torque method for the control of a 2 degrees of freedom orthosis for the rehabilitation of the lower limb, the AIRGAIT exoskeleton's leg orthosis. The actuation of the AIRGAIT is made through self-made pneumatic muscles. For this reason this work starts with the static and dynamic characterization of our pneumatic muscles. The followed approach is based on the analytical description of the system. For this, we describe the pneumatic muscles behaviour with an easy-invertible polynomial fit function in order to model its non-linear trend. We give a geometrical model of the mechanical system to compute the length between the attachments of the pneumatic muscles to the structure for every angles assumed by the two joints. We evaluate through Newton-Euler equation the couples at the joints for each values of the angles. At last we show some validation tests in order to characterize the functioning of the proposed control model on the actuation of the orthosis.

\end{abstract}


\maketitle

\section{Introduction}

Pneumatic artificial muscles (PAMs) are often used for the actuation of rehabilitation devices or, more generally, in most application where there is the interaction between machines and humans \cite{carr2008characterization,kiguchi2008development,zhang2008modeling}.
In these devices, when the motion is not managed by a human, a control model is needed. In literature there are a lot of models for this purpose and applied to PMAs based actuations. The different approaches can be divided into two main groups: feedback linearization and computed-torque method \cite{amato2013robust}. In the first class can be group all the control models that work on the feedback of the measured control variable such as fuzzy \cite{zhang2008modeling}, PID, Neural Network \cite{thanh2006nonlinear} or other models \cite{ariga2012novel,choi2011position}. Many of these control models were tested on 1 degree of freedom systems (\cite{situm2008design,ariga2012novel}) but recently many authors are working on more complex systems that can simulate well the human morphology of the arms or of the legs, then with 2 degrees of freedom, see \cite{thanh2006nonlinear,chang2010adaptive,choi2011position}.

The computed-torque method, instead, requires a complete description of the system and, if it has a high number of degrees of freedom, the formulation of the couple joint expression appears to be very difficult to solve. On the contrary, if the analytical description of the system is well-made, it will be faster to follow the inputs with respect to the other main control model class. In this paper we propose and use a model control based on the computed-torque method for the managing of our AIRGAIT orthosis for the rehabilitation of the lower limb \cite{mat2013design,dzahir2013development,prattico2014couple}. 

The paper is organized as follow. In section \ref{over} we give an overview on the AIRGAIT system. In section \ref{pneu} we show the main characterization of our self-made PAMs. The control model with all its parts is described in section \ref{mode}. Section \ref{vali} contains all the validation tests made on the system in order to verify the goodness of the control model. At last, in section \ref{concl} we give some concluding remarks.

\section{Overview of AIRGAIT exoskeleton's leg orthosis}
\label{over}

\begin{figure}
\centering
\includegraphics[height=8cm]{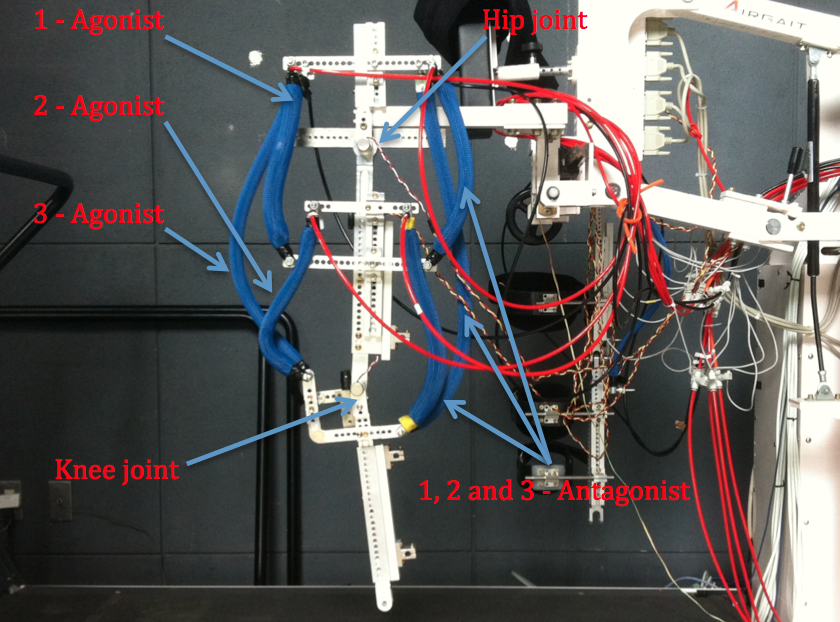}
\caption{AIRGAIT exoskeleton’s leg orthosis}\label{air}
\end{figure}

Figure \ref{air} shows the AIRGAIT exoskeleton’s leg orthosis of the developed body weight support gait training system used for this research. The leg orthosis system implemented six PAM which antagonistically arranged based on the human musculoskeletal system (i.e., mono- and bi-articular muscles). 

The PAM used in this research is a self fabricated McKibben artificial muscle actuator. The input pressure of the PAM is regulated by electro-pneumatic regulator separately for each actuator. The special characteristic of PAM will cause it to contract when the air pressure is supplied, and expand when the air pressure is removed. In other words, the PAM is able to emulate the force and muscle contraction of human’s muscle. In addition, it is also might be able to perform similar contractions and expansions, where their movement is almost similar to the movements of the human’s muscles. The measurement of the joint excursions (i.e., hip and knee) is made using potentiometer. This system uses the Lab-View software and RIO module to provide the input signals and to read the output data of the leg orthosis.

\section{Pneumatic muscle characterization}
\label{pneu}

\begin{figure}
\centering
\includegraphics[height=8cm]{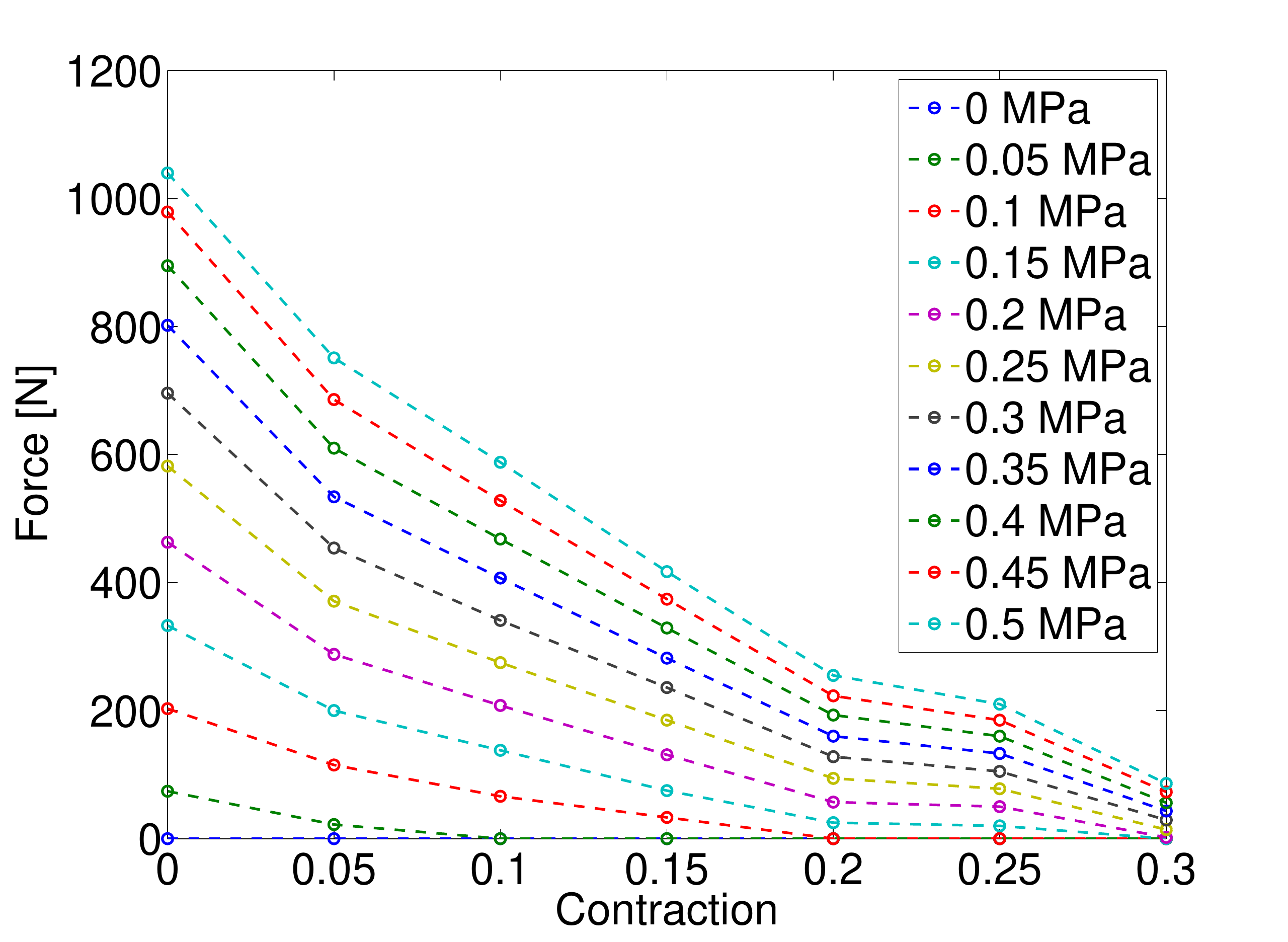}
\caption{Static characterization of the PAM}\label{char}
\end{figure}

The Mckibben PAM used for this study are built in our laboratory using commercial parts. For this reason we have to characterize them in order to understand and fix their properties and behaviours. We conduce two main kind of characterizations, one static and another dynamic. With the data collected by the first one we are able to model the non-linearity of the PAM by fitting the data with a polynomial function. With the dynamic characterization instead, we can estimate $a$ $priori$ the error in position due to the hysteresis. The static characterization is conduced by setting the ends of the PAM at given positions in order to have a variation from 0 to the 30\% of the contraction. This parameter is defined as the difference between the length of the muscle and the given position, divided by the length of the muscle, then:
\begin{equation}
k=\frac{l_m - l}{l_m}
\label{cont}
\end{equation}
Once the distance between the ends is fixed, we vary the pressure supply inside the PAM from 0 to 0.5 $MPa$ and we record, through a load cell, the reaction force. The results of the described experiment are show in figure \ref{char}.
It is possible to note in this figure that the main static properties of the PAM are very similar to those of the commercial PAM.

\begin{figure}
\centering
\subfloat[]{\label{main:n}\includegraphics[height=6cm]{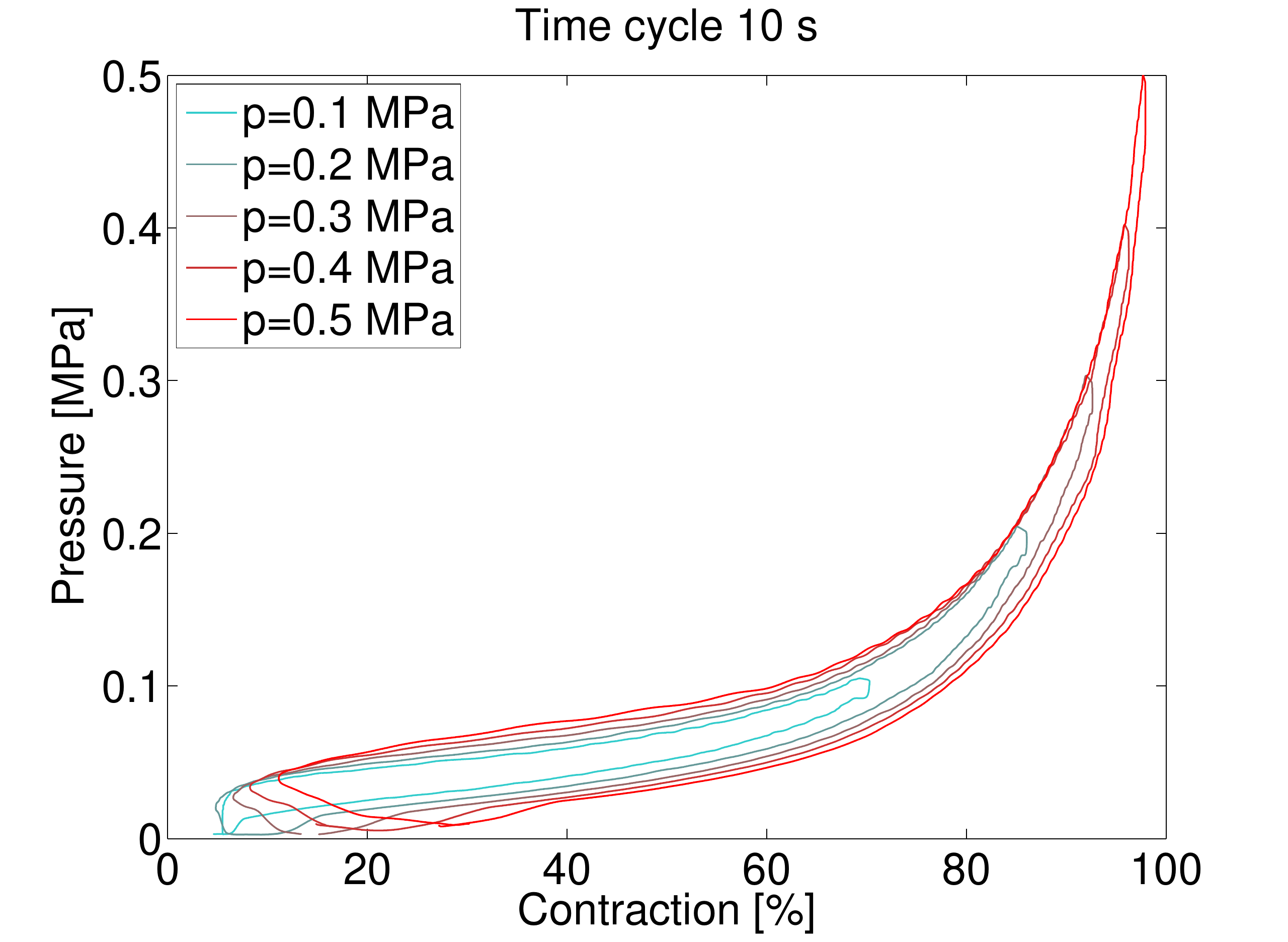}}
\subfloat[]{\label{main:l}\includegraphics[height=6cm]{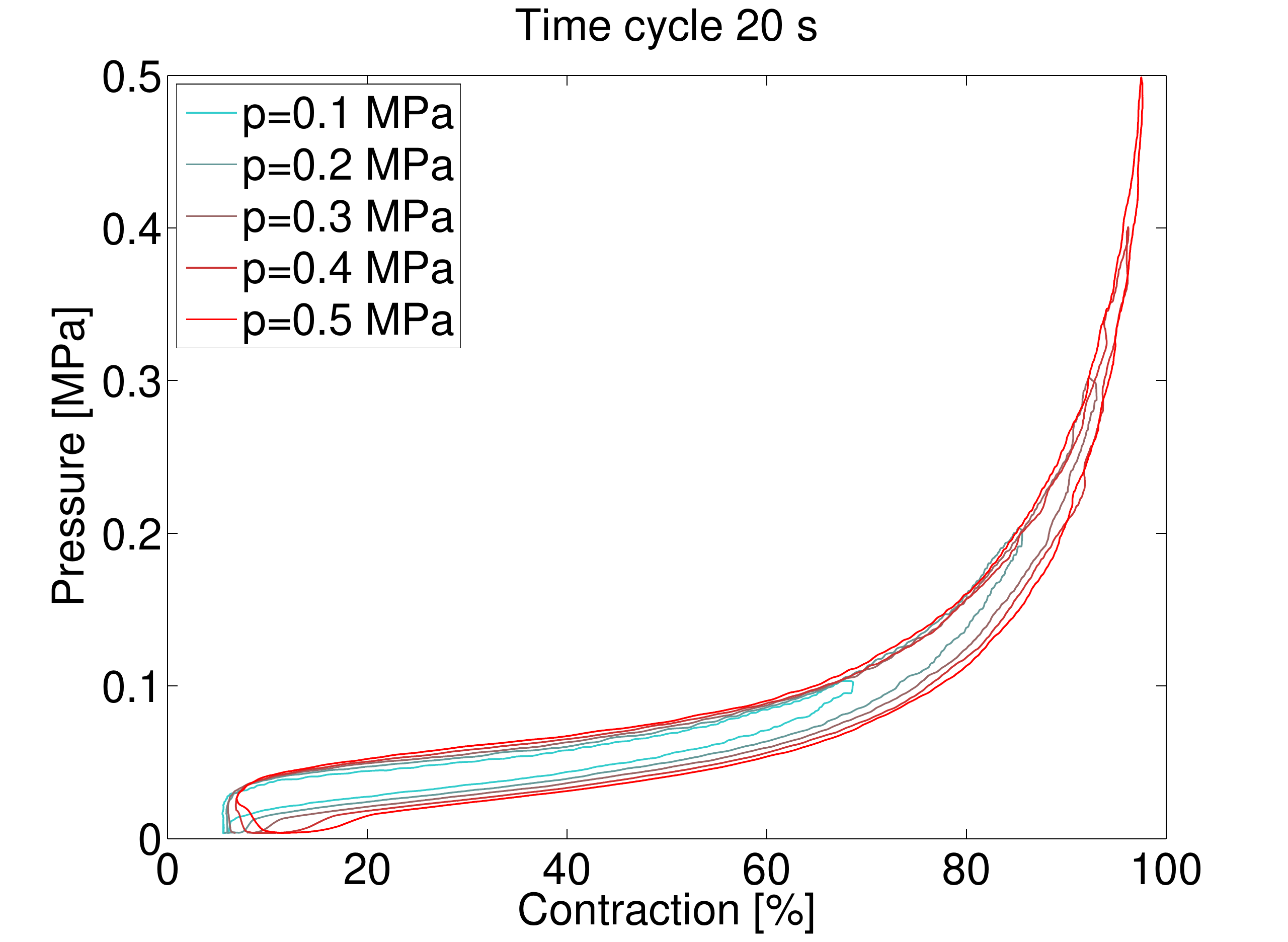}}

\caption{Hysteresis characterization with different time cycle: 10 $s$ (panel $a$) and 20 $s$ (panel $b$).}
\label{hyst1}
\end{figure}

The dynamic characterization allow us to check the ability of the artificial muscle to follow dynamic signals. We conduce two dynamic experiments one with and one without loads. To conduce these experiments we fix the PAM only on one side, maintaining the other free or putting on a weight. We supply the muscle with a pressure signal going from zero to a setted value and once again to zero. The setted we use for the experiments are 0.1, 0.2, 0.3, 0.4 and 0.5 $MPa$. The results of the experiment without loads is presented in figure \ref{hyst1}. The left  panel shows the hysteresis behaviour with a time cycle of 10 $s$. As it is possible to notice, for high values of the pressure 10 $s$ are not enough to complete the loading-unloading cycle. On the right panel, instead, there are the hysteresis trends with a time cycle of 20 $s$. In this case, with all the values of the pressure, the cycle is completed.

\begin{figure}
\centering
\subfloat[]{\label{main:o}\includegraphics[height=6cm]{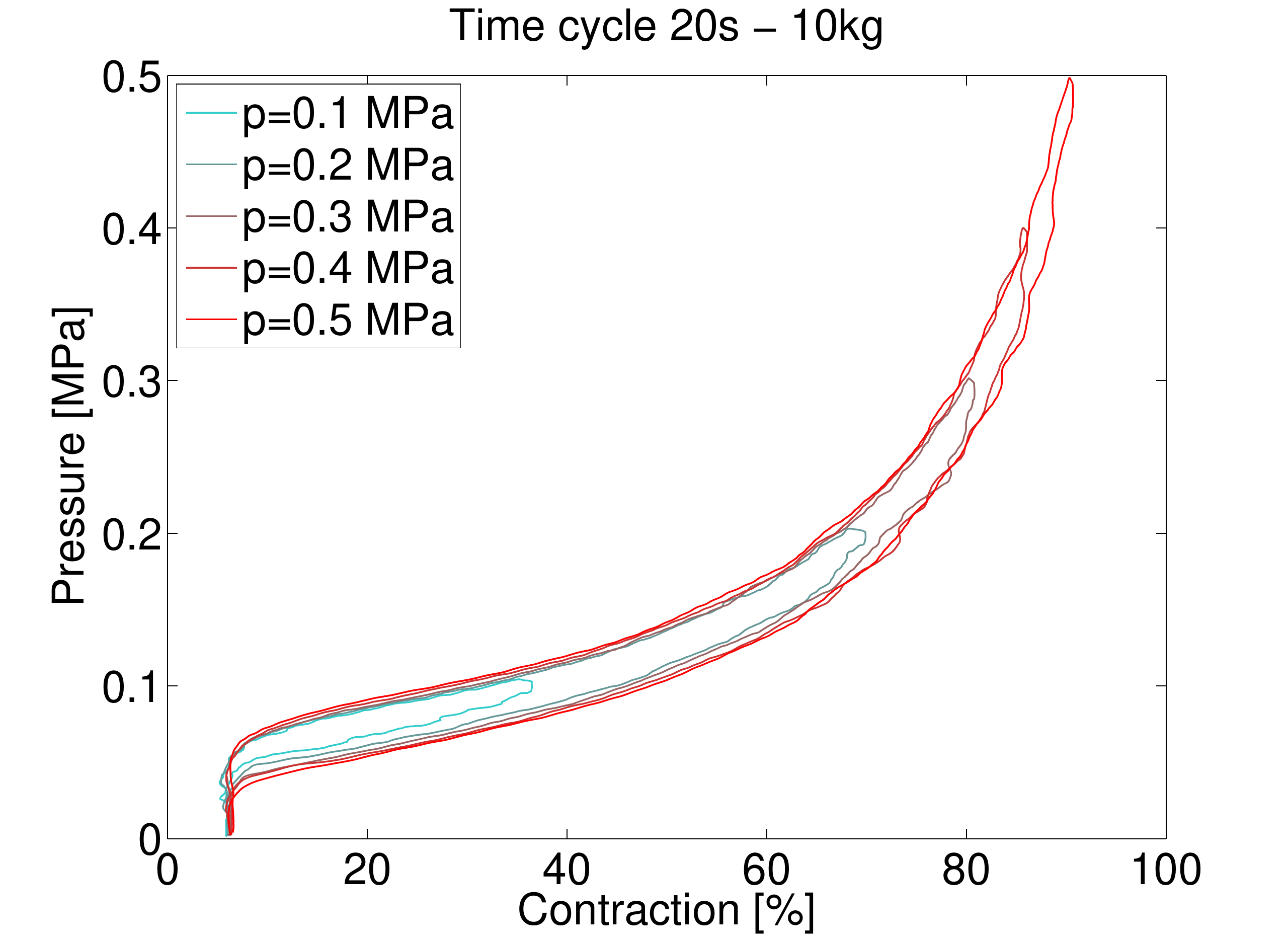}}
\subfloat[]{\label{main:m}\includegraphics[height=6cm]{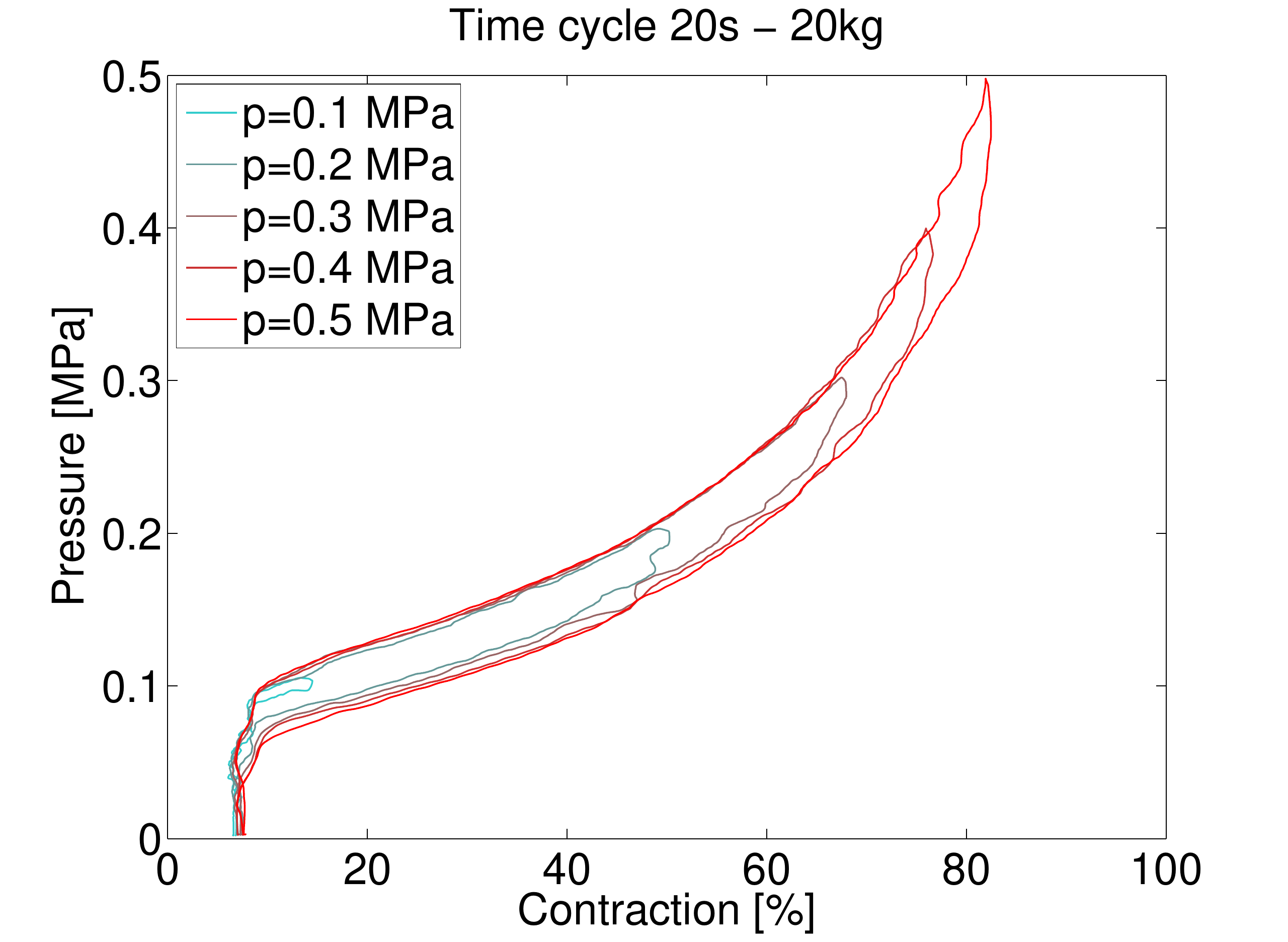}}
\caption{Hysteresis characterization with a time cycle of 20 $s$ and different loads: 10 $kg$ (panel $a$) and 20 $kg$ (panel $b$).}
\label{hyst2}
\end{figure}

Fixing the time cycle to 20 $s$, we conduce the same hysteresis characterization, then loading and unloading cycle, with different maximum pressures, but including a load on the muscle. We test it with 10 and 20 $kg$, that can be considered very high in comparison with the real loads that the system could be stressed. In figure \ref{hyst2}, panel $a$ there is the hysteresis behaviour with a load of 10 $kg$ instead, in panel $b$ that with 20 $kg$. The main interesting consideration can be made by comparing the results of figure \ref{hyst1} with those of figure \ref{hyst2} in terms of distance between the loading and unloading curves. Also with the presence of great load this distance remain almost constant confirming the goodness of these kind of actuation.

\section{Control model and application to the orthosis}
\label{mode}

The control model, proposed in this paper, is based on the analytical description of the system and on the use of the so called computed-torque method. In this section we will show all the main components of the entire control model and the main idea at its basis. 

\subsection{Fitting model of the non-linear behaviour of the PAM}

One of the most difficult problems to solve when we work with PAMs is the non-linear behaviour of the PAMs. 
The main task is to find, as made by \cite{hov2012}, the force that the PAM can apply as a function of the supply pressure and of its contraction. 
 
The data collected into the static characterization (see figure \ref{char}) will be here fit with a surface. We choose to fit the surface with a two variables polynomial function. We need to express the supply pressure as a function of the force and the contraction. To do this, the fitting equation must be solvable in the term of the pressure, then the term of the pressure must have a degree equal or less to two (different approach used in \cite{hov2012} in which the equation is fifth degree in both variables, then needs to solve numerically with long computing time). We then conduce a sensibility analysis on the degree of the fitting equation. Particularly we compute the Root Mean Square Error (RMSE) between the experimental point of Figure \ref{char} and the fitting surface and we express the results as a function of the degrees of the two variables $x$ and $y$ (pressure and contraction). The results are summarized in the Table \ref{st}. As it is possible to notice we have a great reduction of the RMSE from first to second degree in $x$ and, at the same time, we choose to have third degree in $y$. This choice is due to the fact that we do not have a great reduction of the RMSE between third and fourth degree in $y$ and then we decide to reduce the number of the parameters to increase the computational speed.

\begin{table}
\begin{center}

\begin{tabular}{|c|*{3}{c|}|}
     \hline
Degree of $x$ & Degree of $y$ & RMSE [N] \\ \hline
  1 & 1 & 116 \\ \hline
  2 & 1 & 53 \\ \hline
  2 & 2 & 31 \\ \hline
  2 & 3 & 21 \\ \hline
  2 & 4 & 19  \\ \hline
  1 & 2 & 43 \\ \hline
  1 & 3 & 32 \\ \hline
  1 & 4 & 29 \\ \hline

\end{tabular} 
\caption{Sensibility analysis of the fitting curve of the experimental data as a function of the degrees of the polynomial surface}
\label{st} 
\end{center}
\end{table}
The resulting fitting equation is the follow:
\begin{equation}
f(x,y)=a_1 + a_2 x + a_3 y + a_4 x^2 +a_5 x y + a_6 y^2 + a_7 x^2 y + a_8 x y^2 + a_9 y^3
\label{so}
\end{equation}
where, as mentioned before, $x$ represents the supply pressure, $y$ is the contraction and $f(x,y)$ is the force. The numeric values of the parameters of this equation are shown in the Table \ref{par}.

\begin{table}
\begin{center}

\begin{tabular}{|c|*{2}{c|}|}
     \hline
Parameter  & Value \\ \hline
  $a_1$ & -7 \\ \hline
  $a_2$ & 2384 \\ \hline
  $a_3$ & -1135 \\ \hline
  $a_4$ & -467 \\ \hline
  $a_5$ & -12480  \\ \hline
  $a_6$ & 8682 \\ \hline
  $a_7$ & 4160 \\ \hline
  $a_8$ & 13290 \\ \hline
  $a_9$ & -15960 \\ \hline

\end{tabular} 
\caption{Numeric values of the parameters of the fitting polynomial equation}
\label{par} 
\end{center}
\end{table}

\begin{figure}
\centering
\includegraphics[height=9cm]{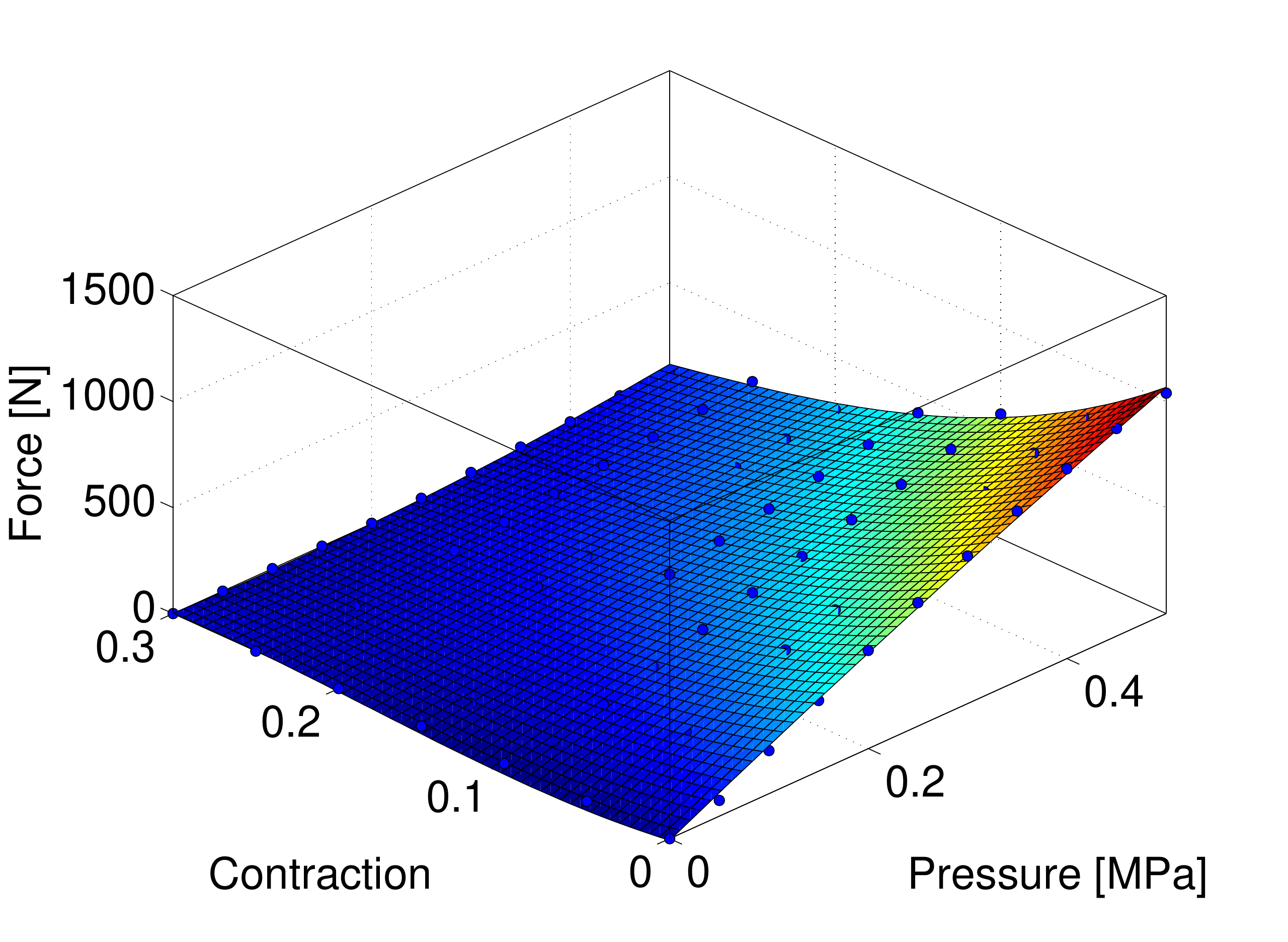}
\caption{Graphic visualization of the fitting polynomial equation.The blue dot are the experimental points}\label{fit}
\end{figure}

At last, we show in figure \ref{fit} the equivalent polynomial surface with the experimental points coming from the characterization. As it is possible to notice from this figure, the equation fits well the real data. 

\subsection{Newton-Euler equation model}

\begin{figure}
\centering
\includegraphics[height=7cm]{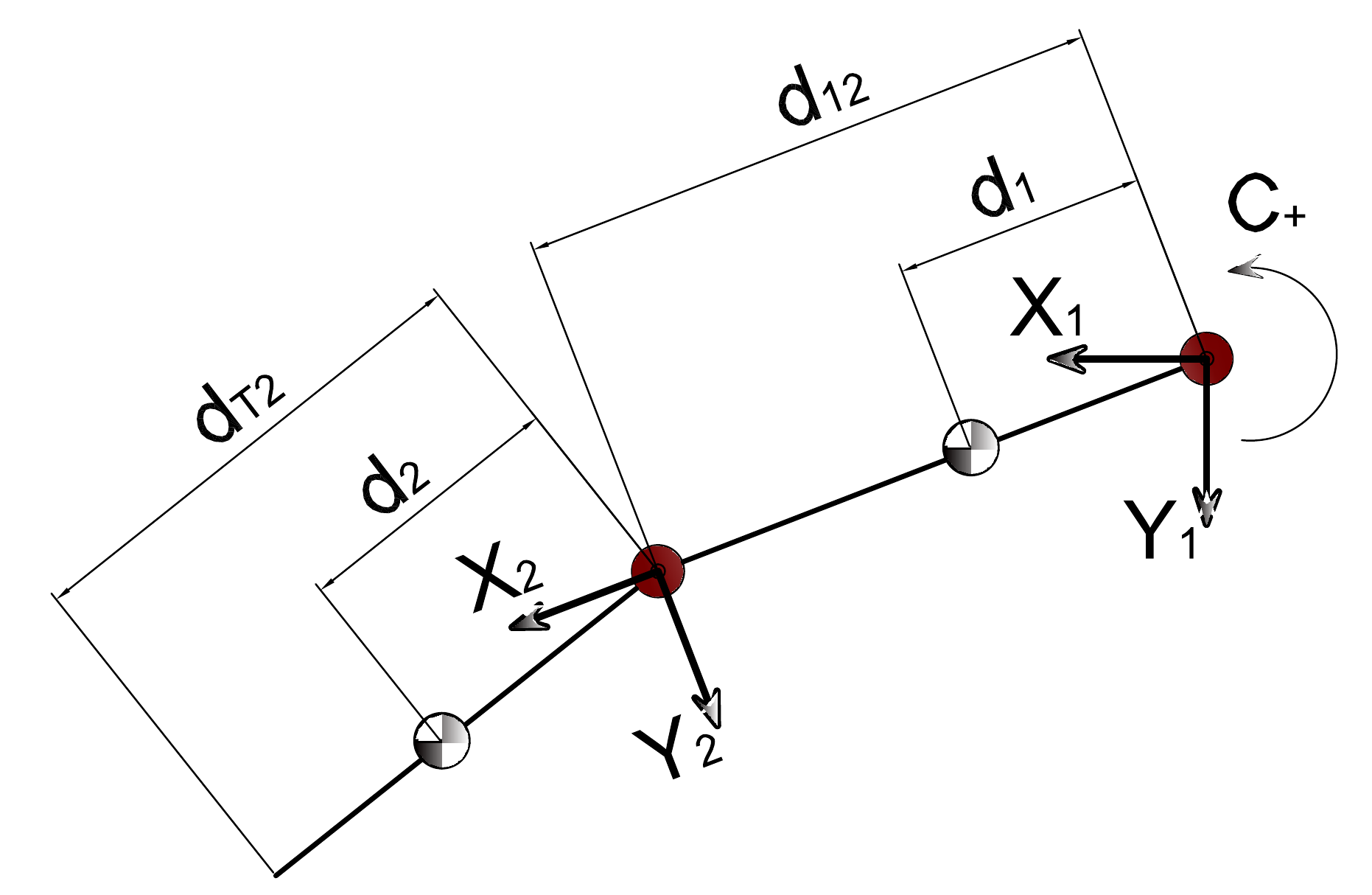}
\caption{}\label{ort}
\end{figure}

The crucial part of the proposed model is based on the computation of the couples for every angles assumed by the two joints. Here we follow the Newton-Euler approach in order to obtain an analytical formulation of the two couples. Just to remind and using a simplified formulation, we can model the dynamics of a robot with revolution joints by the follow equation:
\begin{equation}
M(q)\ddot{q}+C(q,\dot{q})+g(q)=\tau
\label{ne}
\end{equation}

where $\ddot{q}$, $\dot{q}$ and $q$ are respectively the vectors of joint positions, velocities and acceleration, $M(q)$ is the articulated robot inertia matrix, $C(q,\dot{q})$ is the vector of centripetal and Coriolis force, $g(q)$ is the vector of gravitational forces and $\tau$ is the vector of joint torque \cite{amato2013robust}. 
In figure \ref{ort} we give a schematic representation of the orthosis. In this figure $d_1$ and $d_2$ denote the distances between the joints and the centers of mass of the two links instead, $d_{12}$ and $d_{T2}$ are the lengths of the two links.
Referring to figure \ref{ort}, we can solve the equation \ref{ne}, in order to find the couples of the two joints: 
\begin{eqnarray*}
C_1 &=& I_{11} \ddot{\theta_1} + I_{22}\ddot{\theta_1} + I_{22}\ddot{\theta_2} +  \ddot{\theta_1}d_{1}^{2}m_1 + \ddot{\theta_1}d_{2}^{2}m_2 + \ddot{\theta_2}d_{2}^{2}m_2 + \ddot{\theta_1}d_{21}^{2}m_2 + \\
& & d_2 \, g \, m_2 cos( \theta_1+\theta_2) + d_1 \, g \, m_1 cos(\theta_1) + d_{21} \, g \, m_2 cos(\theta_1) - \\
& &\dot{\theta_2}^2 d_2 d_{21} m_2 sin(\theta_2) + 2 \ddot{\theta_1} d_2 d_{21} m_2 cos(\theta_2) + \ddot{\theta_2} d_2 d_{21} m_2 cos(\theta_2) - \\
& & 2 \dot{\theta_1} \dot{\theta_2} d_2 d_{21} m_2 sin(\theta_2)
\label{c1}
\end{eqnarray*}
\begin{eqnarray*}
C_2 &=& I_{22} (\ddot{\theta_1} + \ddot{\theta_2}) + d_2 m_2 ( d_{21} sin(\theta_2 ) \dot{\theta_1}^2 + g\,cos(\theta_1 + \theta_2) + d_2 (\ddot{\theta_1} + \ddot{\theta_2}) +\\
& & \ddot{\theta_1} d_{21} cos(\theta_2) )
\label{c2}
\end{eqnarray*}
where $I$ is the inertia, $m$ is the mass and $g$ is the gravity acceleration. The equations of the two couples are obtained by a symbolic generation of large multibody system dynamic equations proposed in \cite{fisette1996symbolic} and in \cite{marghitu2009mechanisms}. 

In the table \ref{dati} are summarized the numerical data of the orthosis geometry.

\begin{table}
\begin{center}

\begin{tabular}{|c|*{2}{c|}|}
     \hline
Parameter  & Value \\ \hline
  $I_{11}$ & 0.052 $kg m^2$ \\ \hline
  $I_{22}$ & 0.032 $kg m^2$ \\ \hline
  $m_1$ & 1.34 $kg$ \\ \hline
  $m_2$ & 0.97 $kg$ \\ \hline
  $d_1$ & 0.2 $m$  \\ \hline
  $d_2$ & 0.15 $m$ \\ \hline
  $d_{21}$ & 0.4 $m$ \\ \hline
  $d_{T2}$ & 0.37 $m$ \\ \hline

\end{tabular} 
\caption{Numerical data of the orthosis geometry}
\label{dati} 
\end{center}
\end{table}

\subsection{Geometric description model}

\begin{figure}
\centering
\subfloat[]{\label{main:ss}\includegraphics[height=7cm]{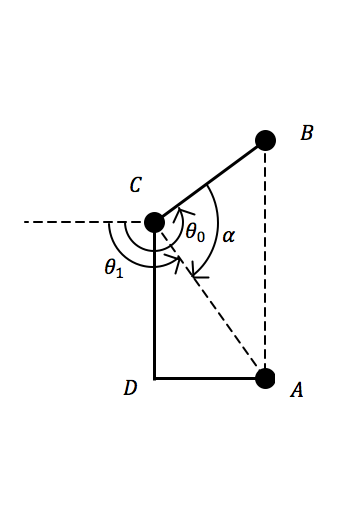}}
\subfloat[]{\label{main:cc}\includegraphics[height=7cm]{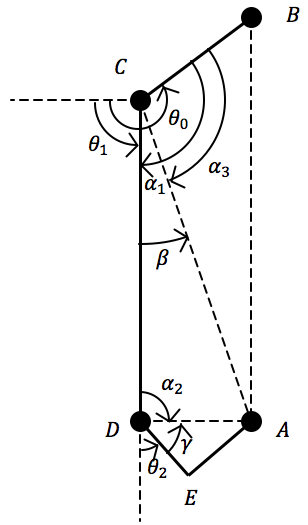}}
\caption{Geometric scheme of the actuation. Panel $a$ mono-articular, panel $b$ bi-articular.}
\label{geo}
\end{figure}

In this section we give the geometric model of the system. We have to describe the variation of the lengths between the ends of the PAMs during the functioning of the orthosis in order to derive the contraction through equation \ref{cont}. Then, we have to find a relation between these lengths $l_i$, related to the muscle $i$, and the joints angles. As the system is made, we have to discern the two cases separately: mono- and bi-articular actuation. These are schematized in figure \ref{geo}, mono-articular in panel $a$ and bi-articular in panel $b$. Another distinction will be made for the two kind of muscle configurations (agonist and antagonist), these due just to the angles coordinate system.

For what concerns the mono-articular configuration (figure \ref{geo}, panel $a$) we can describe the variation of the length of the muscle, defined as $\overline{AB}$, through the use of the law of cosine. Here we show the implementation for the mono-articular hip joint as a function of the angle $\theta_1$, but the same formulation can be derived for the knee joint as a function of the angle $\theta_2$.
\begin{equation}
\overline{AB}=\sqrt{\overline{AC}^2+\overline{BC}^2-2\overline{AC}\cdot\overline{BC}cos(\alpha)}
\end{equation}
where $\alpha$, as mentioned before, will have different value for the two cases of muscle configuration, then:
$$
Agonist: \;\;\;\;\; \alpha=\theta_1 +\theta_0 
$$
$$
Antagonist: \;\;\;\;\; \alpha=\theta_1 -\theta_0
$$
where
$$
\theta_0 =\alpha|_{\theta_1=0}
$$
with these positions we can explicit, through the equation \ref{cont}, the
variation of the contraction as a function of the joint angle:
\begin{equation}
k(\theta_1)=k^{ag}=k^{ant}=\frac{l_m-\overline{AB}}{l_m}
\end{equation}
where $k^{ag}$ and $k^{ant}$ are respectively the contraction of the agonist and antagonist muscles.

The formulation of the contraction of the muscles in the bi-articular actuation, instead, will be related to both angles. Referring to figure \ref{geo}, panel $b$, we can explicit
\begin{equation}
\overline{AC}=\sqrt{\overline{AD}^2+\overline{CD}^2-2\overline{AD}\cdot\overline{CD}cos(\alpha_2)}
\end{equation}
where
$$
\alpha_2=\pi-\gamma-\theta_2
$$
and $\gamma$ is a static angle that can be measure manually on the orthosis. It is equal to 1.89 $rad$ for the agonist side and 0.68 $rad$ for the antagonist one.
\begin{equation}
\overline{AB}=\sqrt{\overline{AC}^2+\overline{CB}^2-2\overline{AC}\cdot\overline{CB}cos(\alpha_3)}
\end{equation}
where
$$
\alpha_3=\alpha_1-\beta
$$
with
\begin{equation}
\beta=acos \left( \frac{\overline{CD}^2 + \overline{AC}^2 - \overline{DA}^2}{2 \overline{CD} \cdot \overline{AC}} \right)
\end{equation}
and distinguishing for the two cases of muscle configurations, $\alpha_1$ is equal to
$$
Agonist: \;\;\;\;\; \alpha_1=\theta_1 +\theta_0 
$$
$$
Antagonist: \;\;\;\;\; \alpha_1=\theta_1 -\theta_0
$$
The contraction for the muscles in the bi-articular actuation can be now evaluated as
\begin{equation}
k(\theta_1,\theta_2)=k^{ag}=k^{ant}=\frac{l_m-\overline{AB}}{l_m}
\end{equation}

\subsection{Control Model}

First of all, we can define the stiffness of a system as the measure of the resistance to the deformations. For our system this concept of stiffness translates itself into the level of the force of the antagonist muscle that we can call the "base force" (following a similar nomenclature proposed by \cite{greb2008}). In order to describe the control model we can set and define, as $R=cost$, the stiffness of the system that represents the force of the PAM that is working against the motion. 

From the geometrical model we can find the contraction of the three pairs of muscles as a function of the angle $\theta_1$ and/or $\theta_2$, then:

\begin{equation}
k_{1}^{ag}=f(\theta_1)  \;\;\; and \;\;\; k_{1}^{ant}=f(\theta_1)
\end{equation}
\begin{equation}
k_{2}^{ag}=f(\theta_2)  \;\;\; and \;\;\; k_{2}^{ant}=f(\theta_2)
\end{equation}
\begin{equation}
k_{3}^{ag}=f(\theta_1,\theta_2)  \;\;\; and \;\;\; k_{3}^{ant}=f(\theta_1,\theta_2)
\end{equation}
where $k_{1}^{ag}$ represents the contraction of the agonist muscle of the joint 1, instead $k_{2}^{ant}$ is the contraction of the antagonist muscle of the joint 2.
From the NE equations we can compute the couples $C_{1}$ and  $C_{2}$ as follow:
\begin{equation}
C_{1}=f(m_1,m_2,I_{11},I_{22},\theta_1,\theta_2,\dot{\theta_1},\dot{\theta_2},\ddot{\theta_1},\ddot{\theta_2})
\end{equation}
\begin{equation}
C_{2}=f(m_2,I_{22},\theta_1,\theta_2,\dot{\theta_1},\dot{\theta_2},\ddot{\theta_1},\ddot{\theta_2})
\end{equation}
but geometrically the couples $C_{1}$ and $C_{2}$ can be also computed as:
\begin{equation}
C_{1}=(F_{1}^{ag}-F_{1}^{ant})\cdot l_i
\label{a1}
\end{equation}

\begin{equation}
C_{2}=(F_{2}^{ag}-F_{2}^{ant})\cdot l_i
\label{a2}
\end{equation}

where $l_i$ is the distance between the $i-th$ muscle force and the joint.
When the orthosis is working, the couples could be both negative and positive. The two cases allow us to distinguish when the agonist or antagonist muscle has to work against the motion and be equal to $R$. For the negative couple case, for example, we can compute the two forces from the equations \ref{a1} and \ref{a2} as follow:
\begin{equation}
F_{1}^{ag}=\frac{C_1}{l}+R
\end{equation}
\begin{equation}
F_{2}^{ag}=\frac{C_2}{l}+R
\end{equation}
The last step of the model consists into solving the fit function of the PAM characterization. Then, using the follow positions
$$
A=a_4 + a_7 y 
$$
$$
B=a_2 +a_5 y+a_8 y^2
$$
$$
C=a_1 + a_3 y +a_6 y^2 + a_9 y^3 -f(x,y)
$$
we can easily solve the equation \ref{so}, as
\begin{equation}
x= \frac{-B \pm  \sqrt{B^2 - 4AC}}{2A}
\end{equation}
and considering the physical meaning of $x$, $y$ and $f(x,y)$, the equation can be summarize as $P=f(F,K)$. Then, known the force and the contraction of the muscle we can compute the required pressure.

\begin{figure}
\centering
\includegraphics[height=4cm]{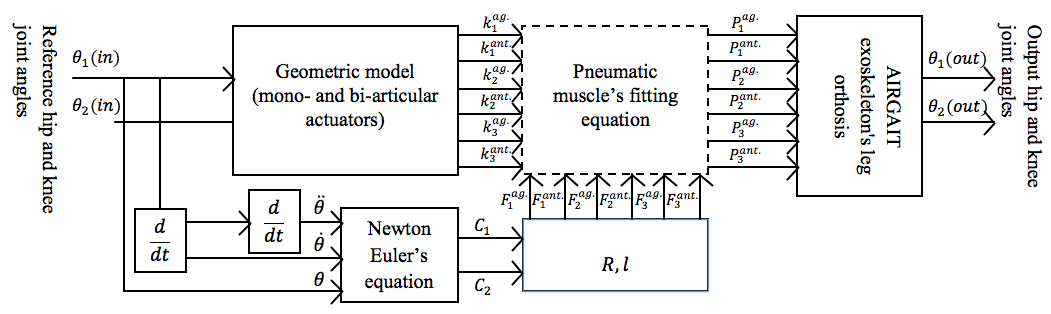}
\caption{Block diagram of the proposed control model}\label{sasa}
\end{figure}

In the figure \ref{sasa} we give the schematic idea of the proposed control model.

\section{Validation tests}
\label{vali}

\begin{figure}
\centering
\subfloat[]{\label{main:s1}\includegraphics[height=6cm]{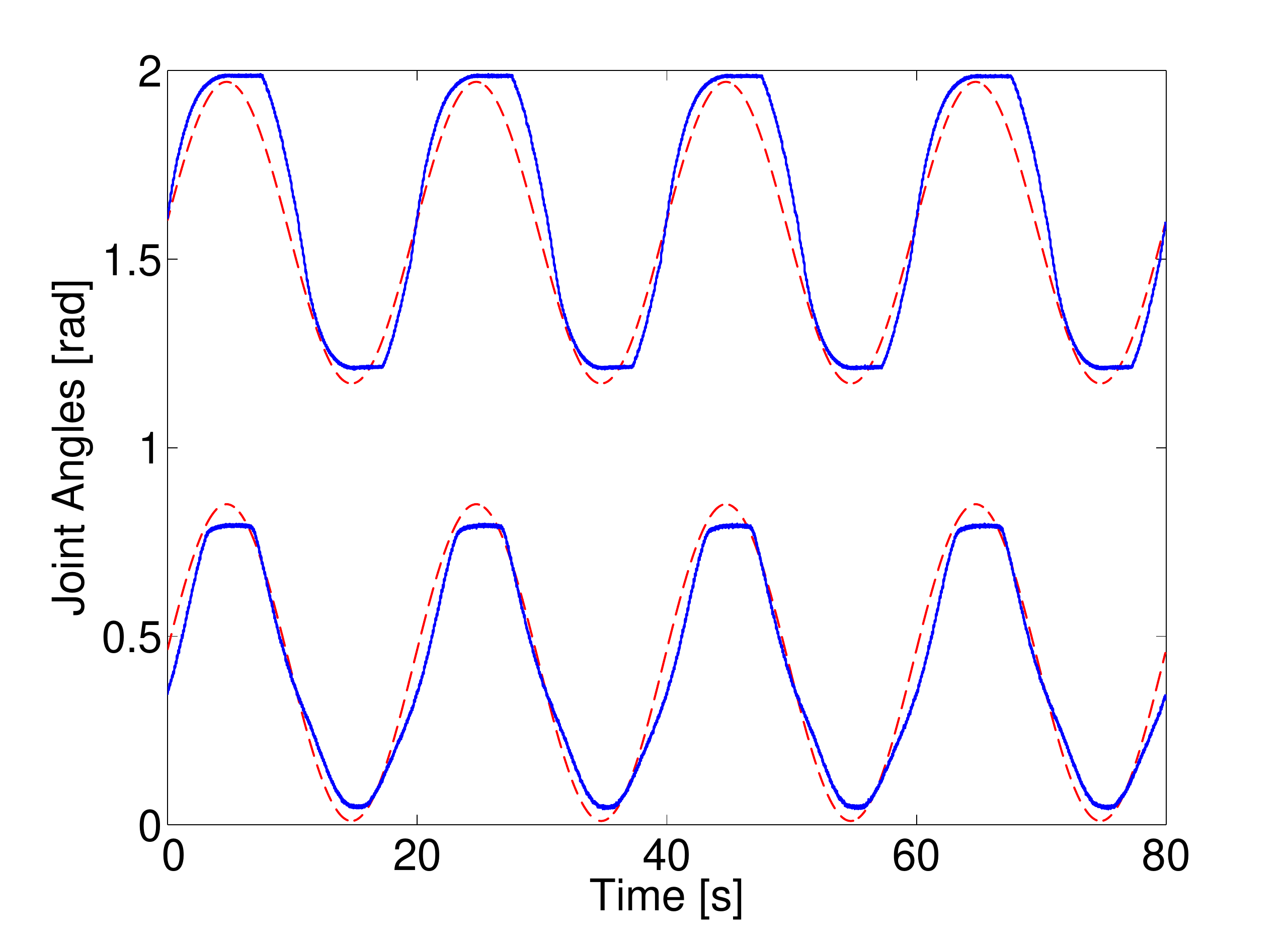}}
\subfloat[]{\label{main:s2}\includegraphics[height=6cm]{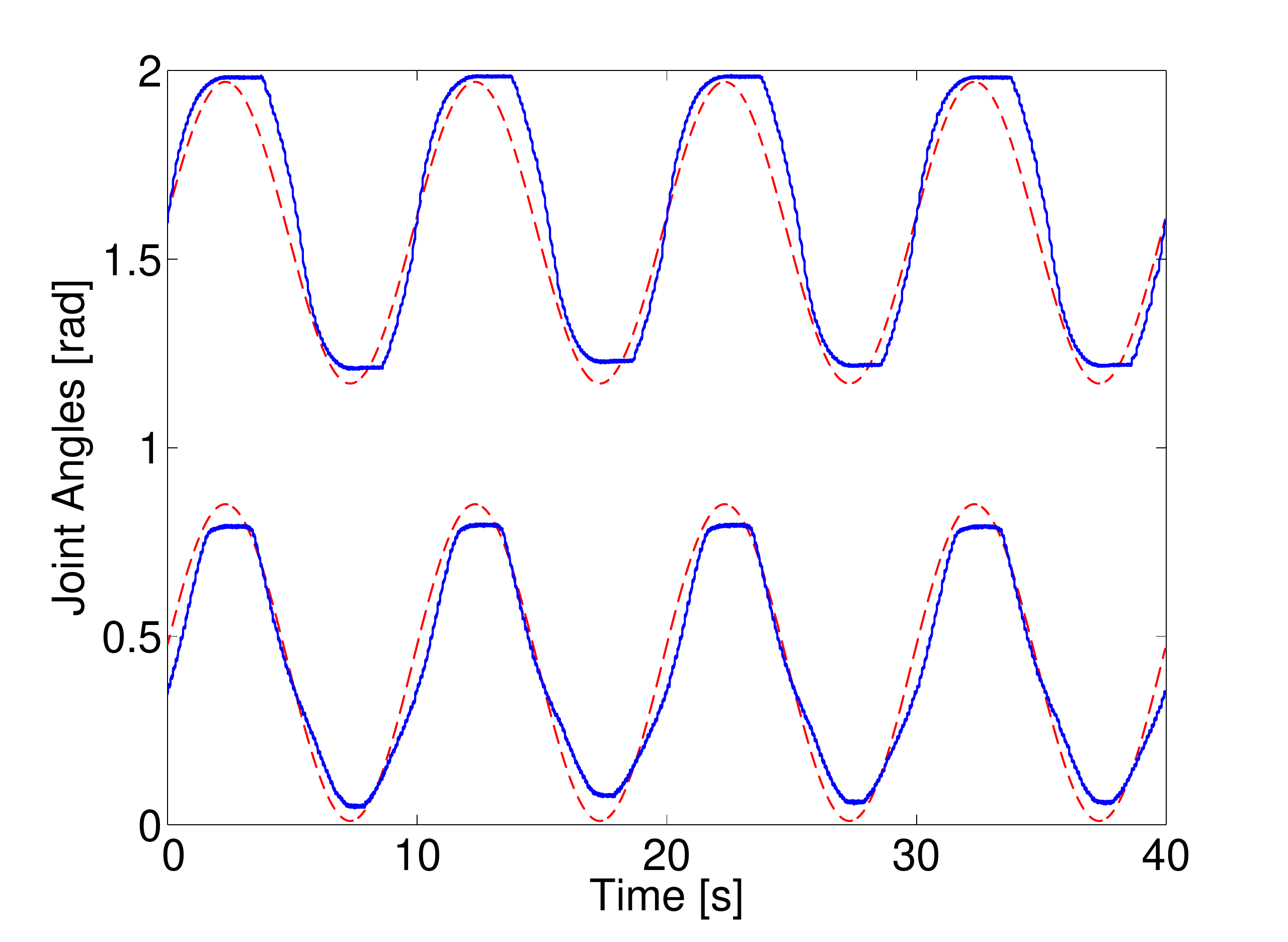}}

\subfloat[]{\label{main:s3}\includegraphics[height=6cm]{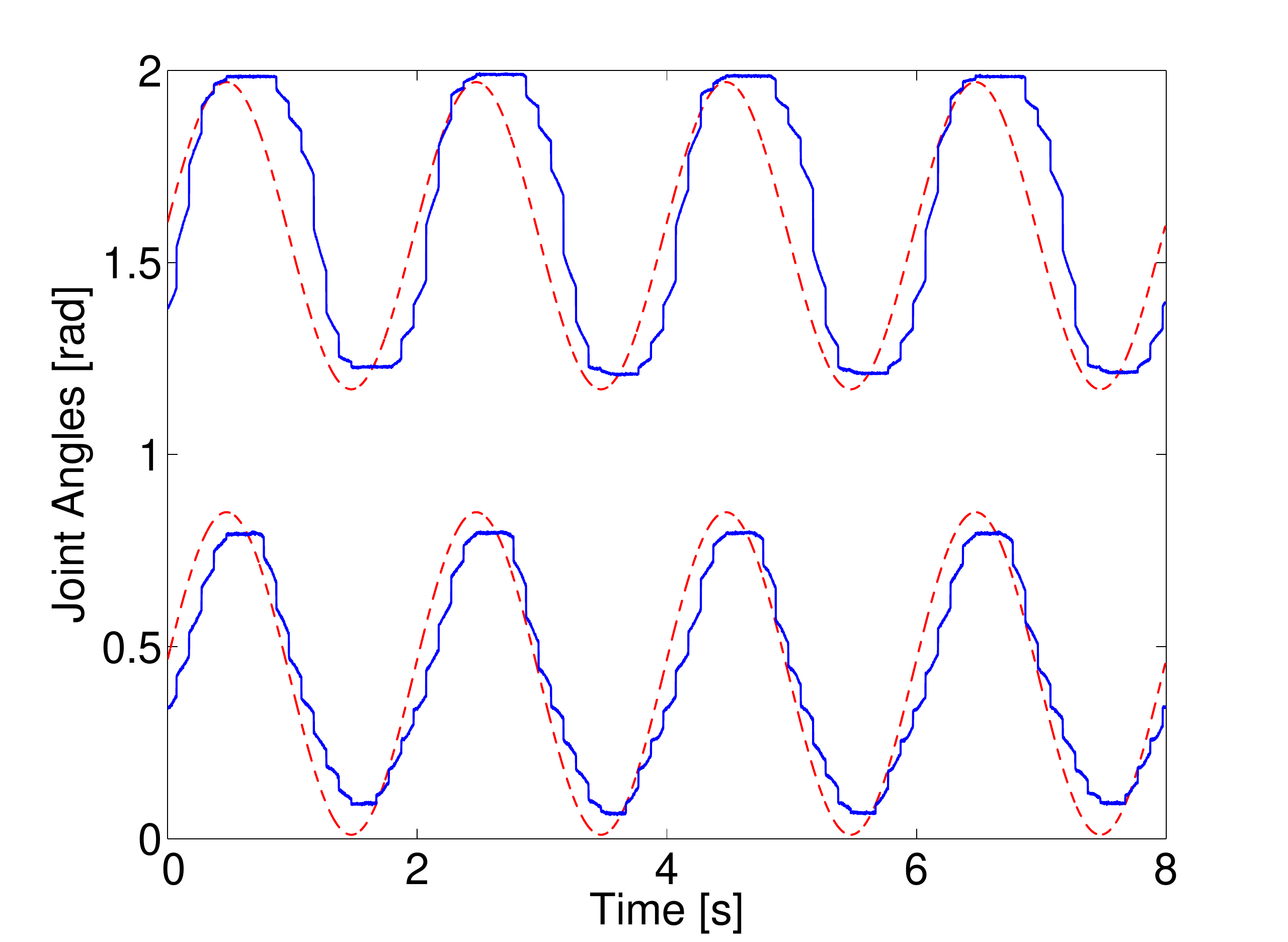}}
\subfloat[]{\label{main:s4}\includegraphics[height=6cm]{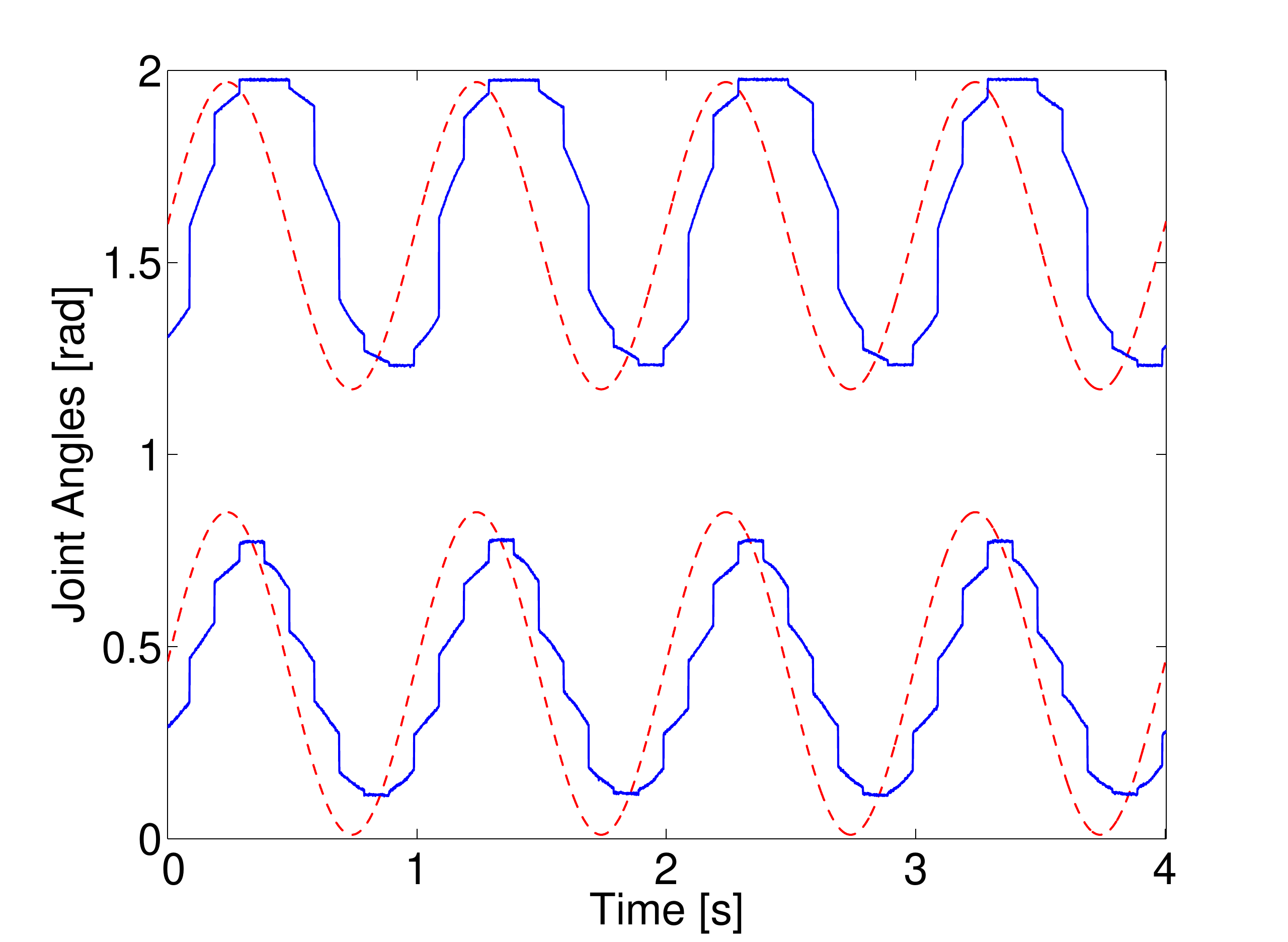}}
\caption{Sine trajectories test for different frequencies. Panel $a$ - 0.05 $Hz$, panel $b$ - 0.1 $Hz$, panel $c$ - 0.5 $Hz$, panel $d$ - 1 $Hz$. The red dashed line is the input signal and the blue continuous line is the measured angles assumed by the orthosis.}
\label{sss}
\end{figure}

In this section we show the results of the validation tests made on the orthosis controlled by the proposed model. We give, as first test, a sinusoidal trajectory to both angles varying its frequency. For the hip joint the sine trajectory has a mean value and an amplitude respectively equal to 1.57 and 0.4 $rad$. Instead, the sine wave, sent to the knee joint, has an amplitude and a mean value both equal to 0.4 $rad$. In figure \ref{sss} are shown the four cases that can be distinguish by the different frequencies of the sine wave that we vary from 0.05 to 1 $Hz$. Particularly here we show the cases of 0.05, 0.1, 0.5 and 1 $Hz$ that corresponds to periods of 20, 10, 2 and 1 $s$. It can be noticed that also in the worst case of a frequency of 1 $Hz$ the system presents a delay but continues to follow almost well the sine wave, with respect to the minimum and maximum values. Moreover we can say that at the frequency of 1 $Hz$ corresponds a walking speed of 1.40 $m/s$ that is the speed of a healthy person \cite{prattico2013new}. Instead for a person that needs of rehabilitation we can consider a speed less or equal to 0.7 $m/s$ at which corresponds a frequency of 0.5 $Hz$. 

\begin{figure}
\centering
\includegraphics[height=8cm]{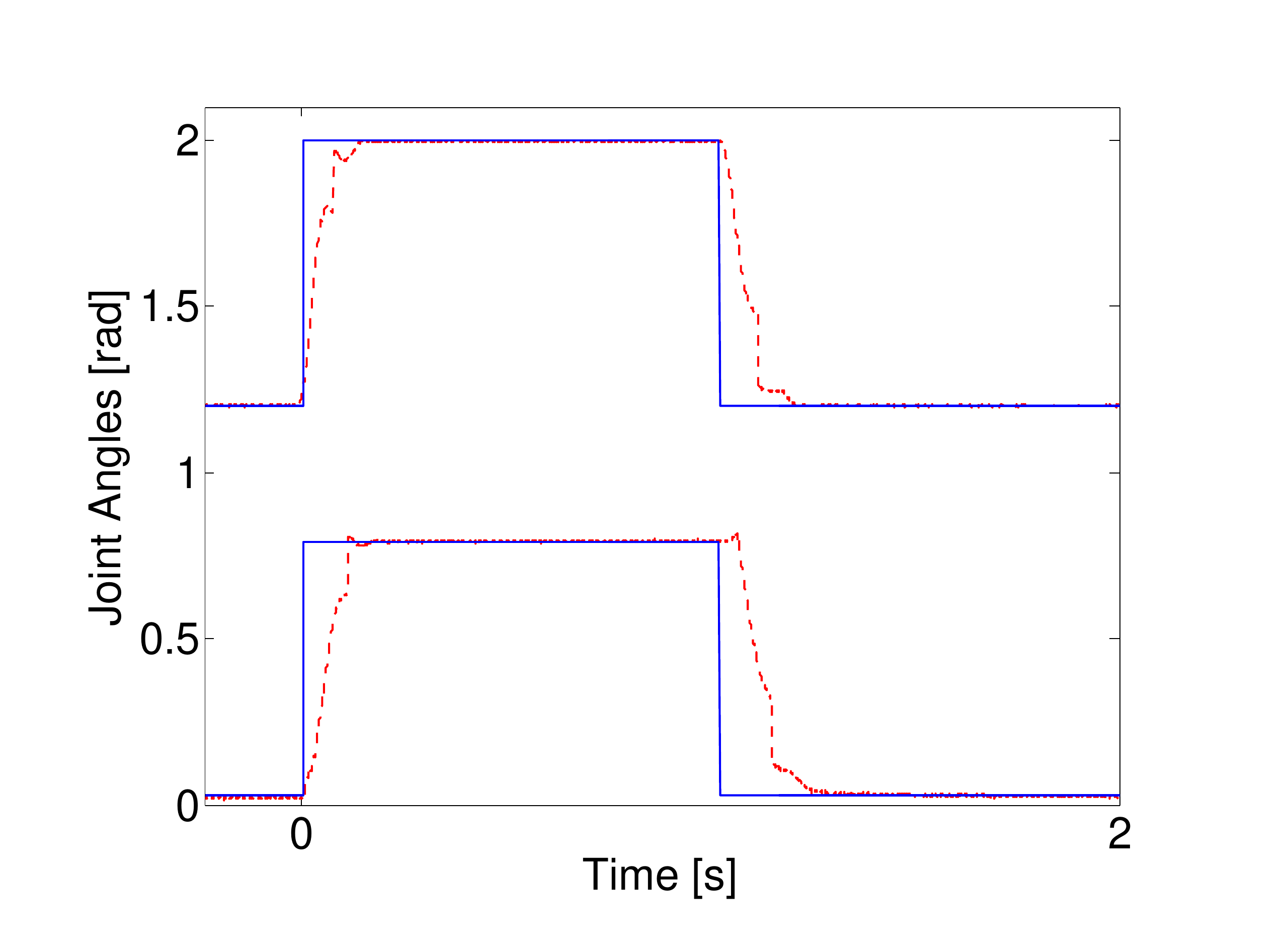}
\caption{Squared trajectory test with a frequency of 0.5 $Hz$. The red dashed line is the input signal and the blue continuous line is the measured angles assumed by the orthosis.}\label{squ}
\end{figure}

Another important test is made by sending a squared signals to both joints. The parameters of the squared trajectories, in terms of mean value and amplitude, are the same of those sinusoidal. Here we just show the case of 0.5 $Hz$. The main scope of stressing the system with a squared wave is to see the response speed. In figure \ref{squ} we show this test and we can noticed that the system is very quick to follow the squared trajectory. Particularly the mean time, considering both the loading and unloading parts, to reach the input signal is equal to 0.1 $s$.

\begin{figure}
\centering
\includegraphics[height=8cm]{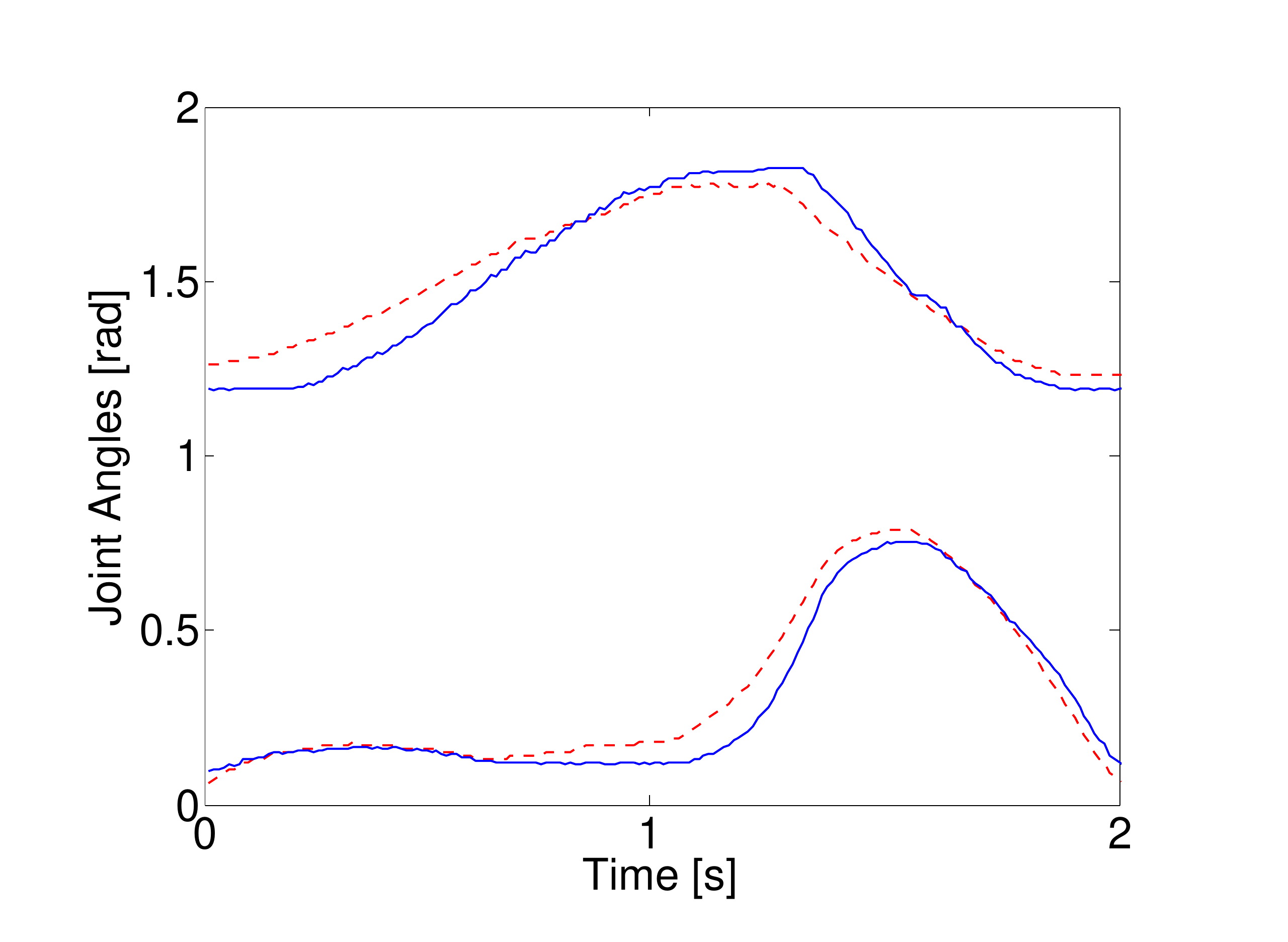}
\caption{Real trajectories for the hip and knee angles for a random walk. The red dashed line is the input signal and the blue continuous line is the measured angles assumed by the orthosis.}\label{walk}
\end{figure}

The last validation test is conduced by recording the hip and knee angles for a random walk and use them as input for the system. By varying the time between the samples we can set easily the cycle speed. Here we show the worst case with a time period of 2 $s$. We can see, from figure \ref{walk} that the input signals are followed with a good accuracy, according to the previous validation tests.

\begin{figure}
\centering
\includegraphics[height=7cm]{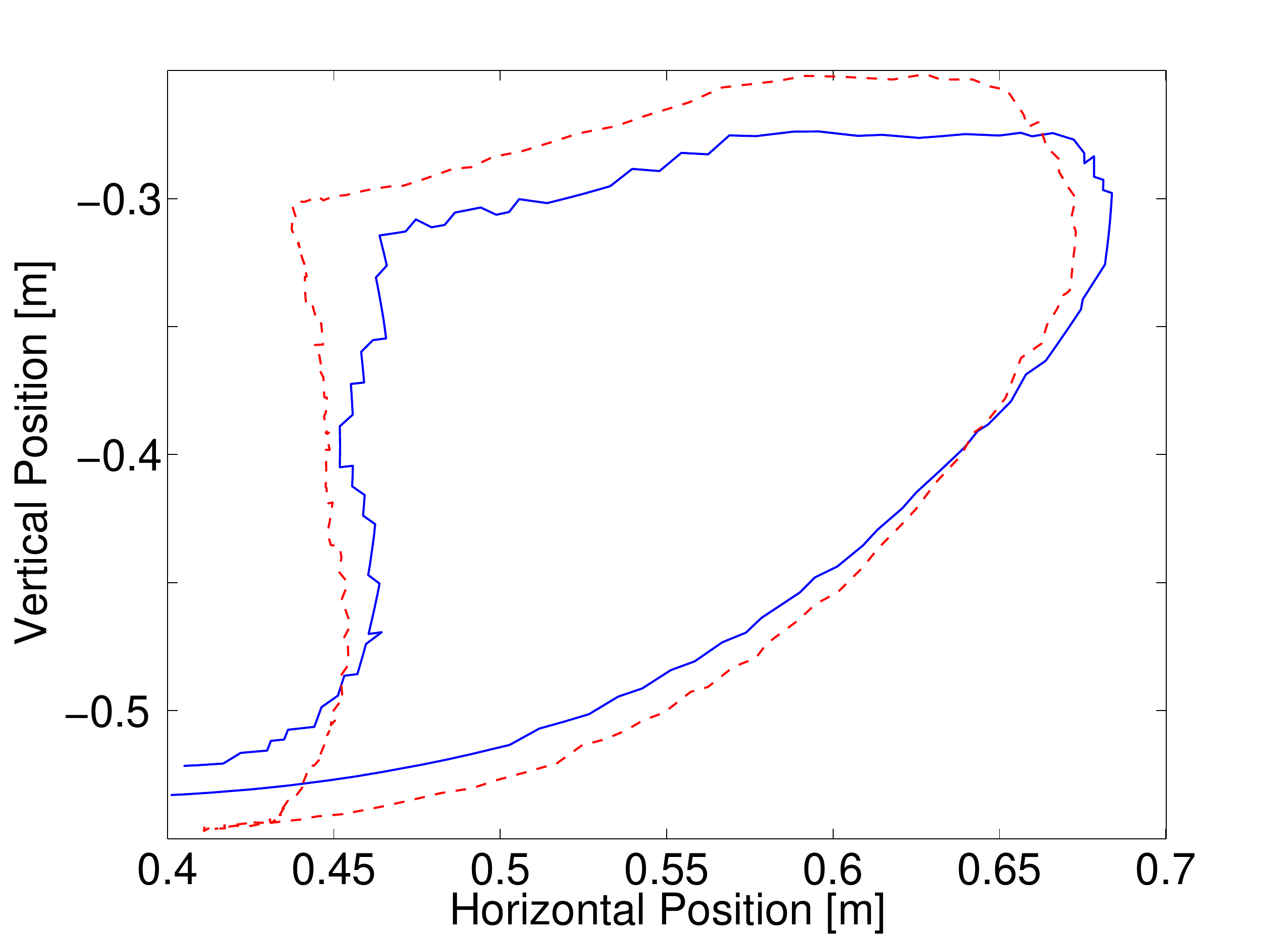}
\caption{Ankle position path for a random walk. The red dashed line is the input signal and the blue continuous line is the real position assumed by the orthosis.}\label{ankle}
\end{figure}

The angles showed in figure \ref{walk} are used in figure \ref{ankle} in order to verify if the system is able to follow a specific path with the end effector, in our case the ankle. We find the position of the ankle just using the equations of the double pendulum, giving the angles of the random walk. We can see in figure \ref{ankle} that the path is well followed.

\section{Discussion and conclusion}
\label{concl}

In this paper we continue the improvement of the control system for our AIRGAIT exoskeleton's leg orthosis. We introduce, with respect to the previous works, the effect of the dynamic components of the system by computing the couples of the joints with the use of the Newto-Euler equations. Moreover, we conduce different validation tests using sine, squared and true random walk trajectories. We show that for the specific purposes for what the orthosis is designed, the PAMs and the proposed control model catch the aim of our work. To the best of our knowledge we are the first on applying the computed-torque method on the control of a two degree of freedom orthosis actuated by PAMs. We show also that, even if there is no managing on the feedback, the proposed model has the advantage to allow the system to follow a given trajectory in a very quickly and with a great accuracy.

\section{Acknowledgement}
This work was supported by KAKENHI: Grant-in-Aid for Scientific Research (B) 21300202. 

\bibliographystyle{elsarticle-harv} 
\bibliography{b}

\end{document}